\documentclass[conference, 9pt]{IEEEtran}
\IEEEoverridecommandlockouts
\IEEEpubid{\begin{minipage}{\textwidth}
    \ \\[45pt]  
    \footnotesize \hspace{-1.8cm}  
    \textcopyright~2025 IEEE. Personal use of this material is permitted. Permission from IEEE must be obtained for all other uses, in any current or future media, including reprinting/republishing this material for advertising or promotional purposes, creating new collective works, for resale or redistribution to servers or lists, or reuse of any copyrighted component of this work in other works. This work is accepted by the IEEE ASRU 2025.
\end{minipage}}
\usepackage{cite}
\usepackage{amsmath,amssymb,amsfonts}
\usepackage{algorithmic}
\usepackage{graphicx}
\usepackage{textcomp}
\usepackage{xcolor}
\usepackage{booktabs} 
\usepackage{enumitem}
\usepackage[hidelinks]{hyperref}
\usepackage{multirow}
\usepackage{tablefootnote}
\usepackage{array} 
\usepackage{amsmath}
\usepackage{amssymb}
\usepackage{makecell}
\usepackage{mathtools}
\usepackage{amsthm}
\usepackage[capitalize,noabbrev]{cleveref}

\def\BibTeX{{\rm B\kern-.05em{\sc i\kern-.025em b}\kern-.08em
    T\kern-.1667em\lower.7ex\hbox{E}\kern-.125emX}}
\begin{document}

\title{GenVC: Self-Supervised Zero-Shot \\ Voice Conversion}

\author{Zexin Cai, Henry Li Xinyuan, Ashi Garg, Leibny Paola Garc\'ia-Perera, Kevin Duh, \\ Sanjeev Khudanpur, Matthew Wiesner, Nicholas Andrews \\
\vspace{.2\baselineskip}
\textit{Human Language Technology Center of Excellence} \\ \textit{Johns Hopkins University} \\
Baltimore, United States \\
\{zcai21, noa\}@jhu.edu
}

\maketitle

\begin{abstract}
\begin{abstract}
Most current zero-shot voice conversion methods rely on externally supervised components, particularly speaker encoders, for training. To explore alternatives that eliminate this dependency, this paper introduces GenVC, a novel framework that disentangles speaker identity and linguistic content from speech signals in a self-supervised manner. GenVC leverages speech tokenizers and an autoregressive, Transformer-based language model as its backbone for speech generation. This design supports large-scale training while enhancing both source speaker privacy protection and target speaker cloning fidelity. Experimental results demonstrate that GenVC achieves notably higher speaker similarity, with naturalness on par with leading zero-shot approaches. Moreover, due to its autoregressive formulation, GenVC introduces flexibility in temporal alignment, reducing the preservation of source prosody and speaker-specific traits, and making it highly effective for voice anonymization.\footnote{Audio samples, code, and model checkpoints are available at \\ \indent\url{https://caizexin.github.io/GenVC/index.html}}
\end{abstract}
\end{abstract}

\begin{IEEEkeywords}
Voice Conversion, Language Model, Speech Anonymization, Speech Synthesis, Speech Generation
\end{IEEEkeywords}

\section{Introduction}
\label{intro}

Zero-shot Voice conversion (VC) seeks to transform a source voice to match an unseen target speaker, with minimal adaptation, while preserving the original linguistic content~\cite{qian2019autovc, zhang20e_interspeech, zhang2022sigvc}. Progress in this area has closely followed advancements in zero-shot text-to-speech (TTS) synthesis, with recent models achieving remarkable naturalness—generating speech that is perceptually indistinguishable from that of real human speakers~\cite{sisman2020overview}. Despite these gains, significant challenges remain, particularly in cloning novel voices and adapting to diverse recording conditions. These limitations arise primarily from the difficulty of training robust and scalable models capable of handling such variability. Furthermore, leading VC approaches operate by converting linguistic features into acoustic representations in a parallel manner. While these systems effectively modify acoustic traits such as timbre, the parallel conversion process often preserves the source speaker’s temporal and prosodic patterns~\cite{mary2006prosodic, li2023freevc, li2024database, cao2024neuralvc}. As a result, the converted speech retains perceptual cues from the original speaker, which diminishes both the naturalness and the similarity of the intended voice transformation~\cite{cai2023identifying, CAI2024privacy}. 

Achieving high-quality zero-shot VC typically requires disentangling speaker identity and linguistic content through two dedicated modules: one that captures vocal characteristics and the other that extracts linguistic content. To achieve this separation, many existing approaches leverage pre-trained models, such as automatic speech recognition (ASR), automatic speaker verification (ASV), or TTS, which rely on supervised training with labeled datasets~\cite{tan2021zeroshot, casanova2022yourtts}. Minimizing the level of supervision in training, however, enhances the scalability of VC systems and allows the exploitation of extensive unlabeled speech corpora~\cite{choi2021neural, choi2023nancy}. One promising strategy for improving zero-shot synthesis involves scaling data to include a broader range of voice types, thereby enhancing generalization to unseen speakers~\cite{betker2023better}. Minimally supervised frameworks like NANCY~\cite{choi2021neural} demonstrate the feasibility of this direction, but often rely on frame-aligned mapping, which can inadvertently preserve source utterance’s temporal and prosodic structure. In contrast, autoregressive architectures provide a compelling alternative to model temporal dependencies more faithfully, and produce prosodic patterns that better reflect the target speaker’s style~\cite{wang2023lm}.

To address these challenges, we propose GenVC, a generative zero-shot VC system designed with three key objectives: (1) reducing dependency on external supervision through self-supervised learning for disentangling speaker and linguistic features, thereby enhancing scalability; (2) leveraging an autoregressive generation mechanism to better model target speaker style and improve voice similarity; and (3) maintaining controllability by encoding speaker characteristics into a compact high-dimensional space, enabling applications such as voice anonymization. To achieve these goals, GenVC employs speech tokenization and is built upon a causal Transformer-based architecture. Experimental results show that GenVC achieves significant improvements in speaker similarity for unseen VC tasks and enhanced privacy preservation in anonymization evaluations, while maintaining naturalness competitive with leading VC methods.

\section{Related Work}
\label{sec:related_works}
VC can be viewed as a subset of speech generation tasks. The rise of self-supervised learning (SSL) models and language models (LMs), known for their strengths in contextual modeling and scalability in natural language processing, has prompted their adoption in sequential audio tasks. Concurrently, neural audio tokenizers, also referred to as speech codecs, have enabled the transformation of continuous audio signals into discrete tokens while maintaining high-fidelity reconstruction. These audio tokenizers are particularly well-suited for integration into LM-based architectures~\cite{kreuk2023audiogen, wu2024towards}, paving the way for a new wave of LM-driven speech generation models. One leading example is AudioLM~\cite{borsos2023audiolm}, which leverages SoundStream~\cite{zeghidour2021soundstream} for tokenization and employs a hierarchical arrangement of three LMs to predict semantic, coarse acoustic, and fine acoustic tokens in sequence. This approach allows for the generation of natural and coherent audio continuations from short prompts. 

These advancements in speech generation have significantly influenced the development of generative zero-shot TTS systems~\cite{kharitonov2023speak, peng2024voicecraft, du2024cosyvoice2, du2025cosyvoice3inthewildspeech, anastassiou2024seedtts}. Among the earliest examples is VALL-E~\cite{microvalle}, which employs Encodec~\cite{fossez2023high} as the audio tokenizer and incorporates both auto-regressive and non-autoregressive LMs to predict audio tokens from input text or phonemes. Later iterations of VALL-E extended its capabilities to tasks such as cross-lingual synthesis and more efficient token representation~\cite{zhang2023speak, chen2024vall}. Beyond purely LM-based models, hybrid systems have also emerged. For example, CosyVoice~\cite{du2024cosyvoice} combines LM-based text-to-token generation with conditional flow-matching models for token-to-speech synthesis.  Another model, MaskGCT~\cite{wang2024maskgct}, extends the LM and tokenization paradigm by replacing the auto-regressive mechanism with non-autoregressive, masking-based generation strategies for efficient audio synthesis~\cite{wang2024maskgct}.

A persistent challenge in LM-based speech generation lies in the nature of current speech codecs, which typically produce multiple tokens—often eight—per frame. This parallel token prediction increases modeling complexity and computational cost. Techniques like delayed token prediction~\cite{copet2023simple, defossez2024moshi} have been proposed to address this issue, yet the overhead remains significant. In contrast, models such as Tortoise-TTS~\cite{betker2023better} and XTTS~\cite{casanova2024xtts} adopt a simpler approach, discretizing audio into a single token per frame. This design choice not only streamlines the model architecture but also aligns more effectively with standard LM training paradigms.

Research on LM-based approaches for zero-shot VC remains limited. Building on the AudioLM framework, Wang et al. proposed LM-VC~\cite{wang2023lm}, which incorporates three LMs to model semantic and acoustic features. While the model demonstrates strong performance in preserving speaker identity and generating natural-sounding speech, its reliance on a multi-component architecture results in slow inference and makes it unsuitable for real-time applications. To address this, StreamVoice~\cite{wang2024streamvoice} introduces a simplified architecture using a single LM to enable streaming zero-shot VC. However, it depends on a separately trained supervised ASR model to extract semantic content. Furthermore, the output speech duration is fixed to match the source speech, preserving prosodic features but also leading to speaker information leakage from the source—an issue shared with many traditional VC systems. 

A concurrent LM-based approach, Vevo~\cite{zhang2025vevo}, addresses similar limitations by introducing a controllable framework for timbre and style conversion. The system comprises two stages: an autoregressive transformer followed by a flow-matching transformer. Both stages are trained with self-supervised, in-context learning, making the framework scalable. However, Vevo conditions speaker representation directly on the source spectrogram rather than using a high-dimensional embedding. While this design is effective for voice imitation, it limits the model’s ability to generate pseudo voices, reducing its suitability for applications like speech anonymization~\cite{tomashenko2024voiceprivacy}.

\begin{figure*}[ht]
  \begin{center}
  \includegraphics[width=0.8\textwidth]{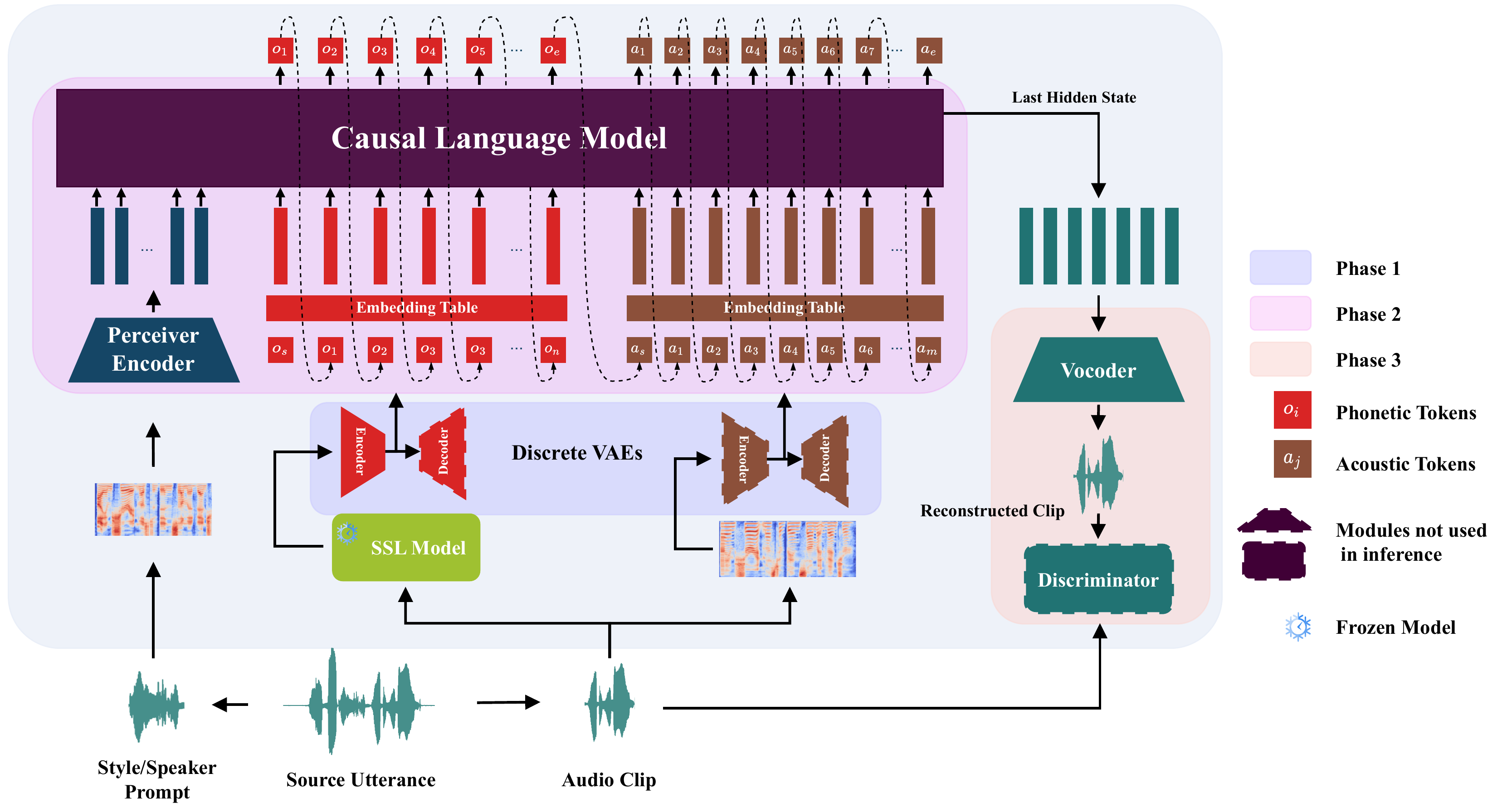}
  \caption{System architecture and training scheme of GenVC:
Phase 1 involves the Discrete VAEs for audio tokenization. Phase 2 has a causal Transformer-based language model alongside a Perceiver encoder. Phase 3 includes a vocoder for waveform reconstruction.}
  \label{fig:genvc}
  \end{center}
  \vskip -0.25in
\end{figure*}


\section{GenVC}
Inspired by prior research in speech generation, we present GenVC, a zero-shot generative VC model. Our approach integrates discrete audio tokenizers with a generative Transformer-based LM for VC, simplifying the complexity of earlier LM-based VC methods. This design enables the model to be trained entirely in an self-supervised manner, supporting large-scale training. Furthermore, the autoregressive nature of the model allows for the conversion of source utterances into the voice and style of a target speaker without preserving the prosodic structure of the source utterances. 

Figure \ref{fig:genvc} presents an overview of GenVC. At its core, GenVC incorporates a causal Transformer-based LM designed to produce acoustic token sequences conditioned on a fixed-length voice style embedding and a sequence of linguistic tokens. Two auxiliary components, the speech tokenizer and vocoder, handle the compression, discretization, and reconstruction of audio. The voice conversion process unfolds in three sequential stages, each building on the last and trained via reconstruction losses, without requiring labeled data or external supervision.

\begin{enumerate}[label=\Roman{*}.]
    \item \textbf{Tokenization}: Modules trained to convert audio signals into discrete tokens, ensuring compatibility with the LM architecture.
    \item \textbf{Generative Modeling}: A decoder-only LM that autoregressively generates audio tokens conditioned on acoustic style representations and linguistic tokens.
    \item \textbf{Vocoding}: A vocoder that reconstructs high-fidelity audio signals from the output of the LM.
\end{enumerate}


\subsection*{Phase 1: Tokenization}
\label{sec:tokenize}
Tokenization is a crucial step in LMs for breaking down inputs into smaller, manageable units. In the context of VC, the objective is to transform the speaker’s timbre/style of a source utterance while preserving its original linguistic content. To achieve this, two distinct types of tokens are used in GenVC: 1. Phonetic Tokens: Capturing the linguistic content of the audio, representing “what” is being said. 2. Acoustic Tokens: Encoding the acoustic properties of the audio, including timbre, prosody, environmental background, and other stylistic elements.

These tokens are derived from different audio representations. Studies have shown that SSL speech models can consistently and significantly produce representations enriched with phonetic information~\cite{choi24b_interspeech}. In the context of VC, SSL models serve as ideal feature extractors for capturing the phonetic units of a given speech signal~\cite{lin21b_interspeech, huang2022s3prl}. For a given input utterance $x \in \mathbb{R}^T$, the phonetic embedding sequence extracted using a pre-trained SSL model is denoted as $\mathbf{O} \in \mathbb{R}^{T_o \times d_o}$, where $T_o$ represents the sequence length, and $d_o$ is the feature dimensionality. For acoustic representations, spectrograms are commonly used to capture the complex properties of audio signals. Such features can be extracted and represented as $\mathbf{A} \in \mathbb{R}^{T_a \times d_a}$, where $T_a$ denotes the sequence length, and $d_a$ specifies the dimensionality of the acoustic features.

As shown in Figure \ref{fig:genvc}, our approach utilizes discrete variational autoencoders (DVAEs)~\cite{van2017neural} to compress $\mathbf{O}$ into phonetic tokens $[\mathbf{o}_1, \dots, \mathbf{o}_n], \mathbf{o}_i \in 1,2,\dots,K_p$ and $\mathbf{A}$ into acoustic tokens $[\mathbf{a}_1, \dots, \mathbf{a}_m], \mathbf{a}_i \in 1,2,\dots,K_a$. Here, $n$ and $m$ denote the lengths of the tokenized sequences derived from the input feature sequences, while $K_o$ and $K_a$ represent the predefined number of discrete codes for the phonetic and acoustic tokenizers, respectively. The DVAE architecture follows those used in Tortoise-TTS~\cite{betker2023better} and XTTS~\cite{casanova2024xtts}.

\subsection*{Phase 2: Generative Modeling}
The backbone model in Phase 2 is an LM built on a decoder-only transformer architecture~\cite{vaswani2017attention}. This phase also incorporates a Perceiver encoder to extract fixed-length style representations from the prompt utterance~\cite{alayrac2022flamingo}. As depicted in Figure \ref{fig:genvc}, the causal LM predicts acoustic tokens by conditioning on both the style prompt derived from the Perceiver encoder and the phonetic tokens extracted from the source audio clip. Note that the audio clip here refers to a segment of the source utterance.

The Perceiver encoder processes a variable-length sequence of acoustic features extracted from the audio prompt and generates a fixed-dimensional audio style representation~\cite{naturalspeech2}. This is achieved by using a set of learned latent vectors as queries, while the keys and values are formed by concatenating these latent vectors with the acoustic features extracted from the prompt audio. The Perceiver encoder architecture consists of cross-attention blocks. The output of the Perceiver encoder $\mathbf{E}_{\text{style}} \in \mathbb{R}^{T_s \times d_\text{model}}$ follows the distribution $P(\mathbf{E}_{\text{style}} \mid \mathbf{A}_\text{prompt}, \mathbf{E}_{\text{latent}}; \theta_\text{Perceiver})$ where $\mathbf{E}_{\text{latent}}$ is the learnable latents that have the same shape as $\mathbf{E}_{\text{style}}$. $T_s$ denotes the fixed number of latent sequence, and $d_\text{model}$ is the dimensionality of the LM. 

In this framework, the model learns the probability distribution $P(\mathbf{o}_{s:e}, \mathbf{a}_{s:e}, \mid \mathbf{E}_{\text{style}};\theta_\text{LM})$ by sequentially estimating the conditional distributions:

\begin{equation}
\label{eq:ploss}
\prod_{t=1}^nP(\mathbf{o}_i \mid \mathbf{o}_{s:i-1},  \mathbf{E}_{\text{style}}; \theta_\text{LLM})
\end{equation}
and
\begin{equation}
\label{eq:aloss}
\prod_{j=1}^mP(\mathbf{a}_j \mid \mathbf{a}_{s:j-1},  \mathbf{o}_{s:e}, \mathbf{E}_{\text{style}}; \theta_\text{LLM}).
\end{equation}

Here, the tokens $\mathbf{o}_s$ and $\mathbf{o}_e$ denote the start and end tokens of the phonetic token sequence, respectively, while $\mathbf{a}_s$ and $\mathbf{a}_e$ mark the start and end of the acoustic token sequence.

The training objective employs two linear prediction heads attached to the final hidden layer of the language model to predict phonetic and acoustic tokens. The overall training loss is designed to maximize the log-likelihood of $P(\mathbf{o}_{s:e}, \mathbf{a}_{s:e}, \mid \mathbf{E}_{\text{style}};\theta_\text{LM})$, as defined in Equation \ref{eq:genloss}. This loss combines the phonetic token classification loss and the acoustic token classification loss, weighted to reflect their respective contributions to the task:

\begin{equation}
    \label{eq:genloss}
    L_{\text{gen}} = \alpha L_{\text{phonetic}} + \beta L_{\text{acoustic}}
\end{equation}

Here, $\alpha$ and $\beta$ denote the weights assigned to the phonetic and acoustic token prediction losses, respectively. The phonetic token loss, $L_{\text{phonetic}}$ is calculated as the negative log-likelihood of the distribution defined in Expression \ref{eq:ploss}, while the acoustic token loss, $L_{\text{acoustic}}$ is the negative log-likelihood of the distribution defined in Expression \ref{eq:aloss}. Since the primary task of the VC system aligns with that of TTS systems—predicting acoustic tokens—we follow the approach outlined in ~\cite{casanova2024xtts} and assign $\beta$ a significantly larger value. 

\subsection*{Phase 3: Vocoding}
Phase 3 focuses on reconstructing the audio waveform from the generative predictions. Since Phase 2 outputs an acoustic token sequence where each token $\mathbf{a}_i$ is highly compressed due to vector quantization, directly reconstructing audio from these acoustic codes often introduces pronunciation issues and artifacts~\cite{casanova2024xtts}. To mitigate these issues, we utilize a HiFiGAN~\cite{NEURIPS2020_c5d73680} vocoder conditioned on features $\mathbf{H} \in \mathbb{R}^{m \times d_\text{model}}$ derived from the final hidden layer of the LM. Unlike systems such as XTTS, which rely on additional speaker embeddings obtained from a separate neural module, we found that conditioning the HiFiGAN vocoder solely on the LM’s final hidden features is sufficient for producing high-quality audio reconstructions.

\subsection{Training and Inference}
During training, as shown in Figure \ref{fig:genvc}, both the audio prompt and the audio clip are extracted from the same source utterance using random start points and segment lengths. Phonetic and acoustic tokens are derived from the audio clip. Typically, the linguistic content of the audio prompt differs from that of the audio clip. This setup allows the Perceiver encoder to capture acoustic attributes that are not represented by the linguistic tokens. Additionally, it facilitates the disentanglement of linguistic content from speaker-specific information in the input audio clip. As a result, the encoder effectively learns to extract both acoustic characteristics and speaker-specific representations from the audio prompt. This approach eliminates the need for supervision from an external speaker embedding model, which is a common feature of traditional zero-shot VC methods. 

During inference, auxiliary components can be omitted, such as the phonetic tokenizer decoder, the entire acoustic tokenizer, and the discriminator. The inference process, given a source utterance $\textbf{Audio}_{\text{src}}$ and a target utterance $\textbf{Audio}_{\text{tgt}}$ specifying the target voice, proceeds as follows:

\begin{itemize}
    \item \textbf{Style Embedding Extraction}: Treat $\textbf{Audio}_{\text{tgt}}$ as the audio prompt. Acoustic features are extracted from the target audio and processed using the Perceiver encoder to obtain the corresponding conditioned style embeddings.
    \item \textbf{Phonetic Token Extraction}: Phonetic features are extracted from $\textbf{Audio}_{\text{src}}$ using the SSL model. These features are then converted into phonetic tokens using the phonetic tokenizer.
    \item \textbf{Acoustic Token Prediction}: The style embeddings and phonetic tokens are fed as input to the backbone LM, which autoregressively predicts the acoustic token sequence until the end token $\mathbf{a}_e$ is generated.
    \item \textbf{Waveform Reconstruction}: The latent representations from the final hidden layer of the LM are passed to the HiFiGAN vocoder to reconstruct the converted audio waveform.
\end{itemize}

\section{Experiments}\label{sec:experiments}
\label{sec:exp}
We conduct a series of experiments to evaluate the effectiveness of our proposed VC approach. Using English speech data, we examine two configurations of the GenVC system: one trained on a smaller dataset and the other on a large-scale dataset. Both subjective and objective evaluations are performed to assess the quality of the converted speech, benchmarking our systems against baseline models. Additionally, we analyze GenVC’s privacy-preserving capabilities and evaluate speaker information preservation from different modules to validate speech disentanglement.

\subsection{Dataset}
Three datasets are used as the training corpus, with their statistics summarized in Table \ref{tab:dataset}. For the LibriTTS dataset~\cite{zen2019libritts}, we include utterances longer than 4 seconds, resulting in a total of 452.5 hours of speech. For large-scale training, we incorporate English utterances exceeding 6 seconds in duration from the CommonVoice~\cite{ardila2019common} and Multilingual LibriSpeech (MLS)~\cite{pratap20_interspeech} datasets. The CommonVoice dataset contributes approximately 1,645 hours of recordings with a diverse range of speakers, while the MLS dataset adds around 44,627 hours of read audiobook speech sourced from LibriVox.

\begin{table}[h]
  \scriptsize
  \caption{Details of the training datasets.}
  \label{tab:dataset}
  \begin{center}
  \begin{sc}
  \vskip -0.1in
  \begin{tabular}[c]{cccccc}
    \toprule
    \textbf{Name}  & \textbf{Dur. (hrs)} & \textbf{\#spk} & \textbf{Avg. length (secs)}  \\
    \midrule
    LibriTTS & 452.5 & 2,278 & 8.59  \\
    CommonVoice-EN  & 1,644.93 & 75,970 &  7.5  \\
    MLS-EN & 44,626.54 & 5,487 & 14.88 \\ 
    \bottomrule
  \end{tabular}
  \end{sc}
  \end{center}
  \vskip -0.1in
\end{table}

For evaluation, we use commonly used test sets in VC and anonymization studies. Specifically, the CMU Arctic~\cite{kominek2003cmu} and Emime~\cite{wester2011emime} datasets are used to construct conversion pairs, enabling the assessment of zero-shot VC performance. Furthermore, in accordance with the speech anonymization pipeline defined by the VoicePrivacy Challenge 2024 (VPC2024)~\cite{tomashenko2024voiceprivacy}, we use the Librispeech dataset~\cite{librispeech} to evaluate linguistic content preservation and privacy protection. We also include VoxCeleb datasets~\cite{nagrani17_interspeech, chung2018voxceleb2} for experiments regarding speaker disentanglement.

\subsection{Training and Inference Details}
\noindent \textbf{Phase 1:} As described in Section~\ref{sec:tokenize}, two DVAE models are trained to tokenize audio into discrete representations. The training data consists of randomly segmented audio clips with a maximum duration of 6 seconds from the LibriTTS dataset. Both models employ a codebook size of 512 and are trained using the Adam optimizer with a learning rate of $1 \times 10^{-4}$. Training is conducted for 200 epochs on an NVIDIA A100 GPU. Each DVAE model contains approximately 52 million parameters.

For phonetic tokens, we use ContentVec~\cite{qian2022contentvec} as the SSL feature extractor. Since ContentVec processes audio signals sampled at 16 kHz, the input audio clips are downsampled accordingly. Features are extracted at a frame rate of 20 ms, and the DVAE further compresses the input sequence length by a factor of 4. Consequently, the phonetic tokens generated by the phonetic DVAE have a token rate of 12.5 Hz. In our experiments, the size of the discrete codebook ($K_o$) for the phonetic DVAE is set to 256. 

For audio tokens, we use Mel-spectrograms as the acoustic features. These are extracted from input audio sampled at 24 kHz, with a window size of 1024 samples and a hop size of 256 samples. The Mel-spectrogram uses 80 Mel bins. This configuration generates acoustic tokens from the acoustic DVAE at an approximate token rate of 23.44 Hz. The size of the discrete codebook ($K_a$) for the acoustic DVAE is set to 1024.

\vskip 0.03in
\noindent \textbf{Phase 2:} The Perceiver encoder is configured with 32 latent queries. We use GPT-2~\cite{radford2019language} as our LM. For the GPT-2 model, the model dimension is set to 1024, with 30 layers. For the input audio prompts, the segments are randomly chosen within a duration range of 3 to 6 seconds. The audio clips used for phonetic token extraction and acoustic token prediction are also randomly cut from the source utterances, with durations ranging from 1.2 to 8 seconds.

We use the Adam optimizer with an initial learning rate of $1 \times 10^{-4}$, and employ a multistep learning rate scheduler, reducing the learning rate by a factor of 0.5 every 5 epochs during training. For loss calculation, the weights $\alpha$ and $\beta$ in Equation \ref{eq:genloss} are set to 0.01 and 1, respectively. The LM consists of 423.64 million parameters, and the Perceiver encoder contains 8.48 million parameters.

For causal LM inference, the temperature is set to 0.85, with a length penalty of 1.0 and a repetition penalty of 2.0. The top-\textit{k} and top-\textit{p} sampling parameters are set to 15 and 0.85, respectively.

\vskip 0.03in
\noindent \textbf{Phase 3:} The vocoder is trained on audio chunks with a fixed duration of 0.64 seconds, producing audio at a sample rate of 24 kHz. As the acoustic DVAE downsamples the sequence, reducing the original acoustic feature length by a factor of 4, the features from the final hidden layer of the LM are upsampled and interpolated by a scale factor of 4 during both training and inference. These interpolated features are then used as the input for vocoder training. 

The vocoder’s generator and discriminator are optimized using the AdamW algorithm with a learning rate of $2 \times 10^{-4}$. The HiFiGAN generator contains 3.16M parameters. Additionally, four types of discriminators are employed for vocoder training: multi-scale discriminator, multi-period discriminator, multi-scale STFT discriminator, and multi-scale sub-band constant-Q transform discriminator~\cite{gumultisubband}.
\vskip 0.03in
\noindent \textbf{Models} We trained two models, GenVC-Small and GenVC-Large, for this study. Both models use the same audio tokenizers trained on the LibriTTS dataset but differ primarily in the training data used for Phase 2 and Phase 3. 

\textbf{GenVC-Small} is trained on the LibriTTS dataset, which comprises approximately 450 hours of audio. Phase 2 training was conducted with a batch size of 32 for 590k steps, followed by Phase 3 vocoder training with a batch size of 64 for about 1M steps. 

\textbf{GenVC-Large} is fine-tuned and initialized from GenVC-Small using additional data from the CommonVoice-EN and MLS-EN datasets. This training was performed with the Perceiver encoder kept frozen. The language model was fine-tuned with a batch size of 24 for 1.5M steps, while the HiFiGAN vocoder was trained with a batch size of 128 for 220k steps.

\section{Results}\label{sec:results}
\subsection{Conversion Performance}
We evaluate the performance of our proposed systems using conversion pairs constructed from the test set. For comparison, we include three leading zero-shot systems—YourTTS, FreeVC, and Neural VC—since their code and pretrained checkpoints are publicly available for inference. Similar approaches, such as LM-VC~\cite{wang2023lm} and StreamVoice~\cite{wang2024streamvoice}, are excluded due to the lack of released implementations. Note that the selected baseline systems were trained on the VCTK dataset and rely on an external ASV system to render speaker identity.

\subsubsection{Objective Evaluation} We use open-source SPV and MOS prediction models to objectively evaluate the speaker similarity and naturalness of our proposed system in comparison to the baseline systems. We randomly construct 2,000 conversion pairs from the EMIME dataset, which serves as a test set with unseen speakers for all VC models. Speaker similarity is measured using a WavLM-based speaker verification system~\cite{chen2022wavlm}.\footnote{\url{https://huggingface.co/microsoft/wavlm-base-sv}} For each pair, we compute the cosine similarity between the embeddings of the target speaker’s utterance and its corresponding converted utterance. The final score is reported as the average cosine similarity across all test pairs. Naturalness is assessed using UTMOS$_\text{v2}$~\cite{baba2024utmosv2},\footnote{\url{https://github.com/sarulab-speech/UTMOSv2}}, which predicts mean opinion scores (MOS) for the converted speech. Higher values indicate better performance for both metrics.

The results are presented in Table \ref{table:obj_eva}. YourTTS achieves the highest speaker similarity, closely followed by Neural VC and FreeVC. GenVC-Large surpasses FreeVC and Neural VC, while GenVC-Small remains competitive. Overall, similarity scores are close among all systems. For naturalness, FreeVC achieves the best performance, while GenVC-Small and GenVC-Large slightly outperform Neural VC and significantly exceed YourTTS. 

\begin{table}[!h]
    \footnotesize
    \caption{Objective evaluation results. UTMOS are reported with a 95\% confidence interval.}
    \label{table:obj_eva}
    
    \begin{center}
    \begin{sc}
    \vskip -0.1in
    \begin{tabular}[c]{@{\ \ \ }l@{\ \ \ \ \ \ }cc@{\ \ \ }}
        \toprule
        \textbf{System}  & \textbf{SIM} $\uparrow$ & \textbf{UTMOS} $\uparrow$\\
        \midrule
        YourTTS~\cite{casanova2022yourtts} & \textbf{0.885} & 2.43 $\pm$ 0.018 \\
        FreeVC~\cite{li2023freevc} & 0.878 & \textbf{2.81 $\pm$ 0.014} \\
        Neural VC~\cite{cao2024neuralvc} & 0.881 & 2.60 $\pm$ 0.015\\
        \midrule
        GenVC-Small & 0.869 & 2.66 $\pm$ 0.021\\ 
        GenVC-Large &0.884 & 2.65 $\pm$ 0.019 \\ 
        \bottomrule
    \end{tabular}
    \end{sc}
    \end{center}
    \vskip -0.1in
\end{table}

\subsubsection{Subjective Evaluation} For the subjective evaluation test, we randomly selected 50 pairs of utterances for conversion. A total of 13 participants were recruited to perform a listening test, in which they rated the naturalness and speaker similarity of genuine and voice-converted utterances. Ratings were given on a scale from 1 (Bad) to 5 (Excellent), with 0.5-point increments. Each participant rated a total of 330 utterances: 250 utterances from five VC systems, 80 genuine audio samples. The stimuli are presented in random order, and detailed instructions are provided on the \href{https://caizexin.github.io/GenVC/index.html}{sample webpage}.

The naturalness mean opinion score (NMOS) and similarity mean opinion score (SMOS) results are summarized in Table \ref{table:sub_eva}. We use the Mann-Whitney U test to assess the statistical significance of all possible pairwise comparisons between different VC approaches~\cite{rosenberg17_interspeech, cooper2025good}. Most pairwise differences in MOS are statistically significant (\textit{p}-value $< 0.05$), except for the NMOS and SMOS comparisons between NeutralVC and FreeVC, and the SMOS comparison between FreeVC and GenVC-Large.

\begin{table}[!h]
    \footnotesize
    \caption{Subjective evaluation results on unseen voice conversion. All MOS are reported with a 95\% confidence interval.}
    \label{table:sub_eva}
    \begin{center}
    \begin{sc}
    \vskip -0.1in
    \begin{tabular}[c]{@{\ \ \ }l@{\ \ \ \ \ \ }cc@{\ \ \ }}
        \toprule
        \textbf{System}  & \textbf{NMOS} $\uparrow$ & \textbf{SMOS} $\uparrow$ \\
        \midrule
        VCTK & 4.23 $\pm$ 0.12 & - \\
        Train-Small & 4.15 $\pm$ 0.14 & - \\
        Train-Large & 4.04 $\pm$ 0.08 & - \\
        \midrule
        YourTTS & 2.96 $\pm$ 0.09 & 2.87 $\pm$ 0.07 \\
        FreeVC & \textbf{3.96 $\pm$ 0.07} & 3.48 $\pm$ 0.07 \\
        Neural VC & 3.89 $\pm$ 0.08 & 3.39 $\pm$ 0.07 \\
        \midrule
        GenVC-Small & 3.57 $\pm$ 0.09 & \textbf{3.72 $\pm$ 0.08}\\ 
        GenVC-Large & 3.43 $\pm$ 0.09 & 3.50 $\pm$ 0.08 \\ 
        \bottomrule
    \end{tabular}
    \end{sc}
    \end{center}
    \vskip -0.1in
\end{table}

According to the results, FreeVC achieves the highest NMOS of 3.96, indicating the best perceived naturalness among the evaluated VC systems. In comparison, GenVC-Small and GenVC-Large achieve lower NMOS scores of 3.57 and 3.43, respectively. However, the quality of training data might impact perceived naturalness. While the baseline systems were trained on the high-quality VCTK dataset, GenVC was trained on datasets with comparatively lower quality, which may have contributed to the observed differences in NMOS.

Our proposed VC approach demonstrates strong performance in speaker similarity. GenVC-Small achieves the highest SMOS of 3.72, surpassing all other systems, including the best-performing baseline, FreeVC, which scores 3.48. However, GenVC-Large does not show further improvement in similarity despite being trained on a larger dataset, though it still outperforms the baseline systems. This may be partially attributed to the quality of the additional training data. Overall, the results suggest that the GenVC models achieve state-of-the-art speaker similarity while maintaining competitive naturalness. Nevertheless, a key limitation remains: although the model’s self-supervised nature supports large-scale training, incorporating larger but lower-quality datasets can negatively affect conversion quality.

\subsection{Anonymization Performance}
Although VC can effectively anonymize voices in terms of timbre, many VC-based approaches reveal source speaker information through the prosodic structure of the speech, thereby limiting privacy preservation. Here we adopt the anonymization and evaluation pipeline from VPC2024~\cite{tomashenko2024voiceprivacy} to assess our models’ anonymization performance in comparison with other VC-based approaches. 

The privacy evaluation pipeline follows a standard speaker verification process. In this setup, the verification model is trained on anonymized data labeled with the original speaker identities. A successful anonymization system should sufficiently distort and obscure the original speaker’s identity, making it difficult for the verification system to accurately identify the source speaker. During evaluation, pairs of source speech, referred to as trials, from the evaluation dataset are anonymized and treated as both enrollment and test speech. The primary metric for privacy evaluation is the equal error rate (EER), calculated based on similarity scores from these pairs. A lower EER indicates a higher risk of speaker re-identification, while a higher EER reflects better performance in preserving voice privacy. In addition, utility evaluation is performed to measure the preservation of linguistic content. The anonymized utterances are transcribed using a speech recognition system, and the resulting transcripts are compared with the ground-truth content from the source data. The word error rate (WER) is used as the metric, with a lower WER indicating better preservation of the linguistic content.

For this experiment, the verification models are trained on converted utterances from the train-clean-360 set of Librispeech and tested on the development and test sets of Librispeech. Target voices are randomly selected from EMIME. As shown in Table \ref{tab:anon}, other VC-based anonymization approaches present risks of revealing the source speaker’s identity, as their average EERs are close to or below 12\%. YourTTS, FreeVC, and Neural VC demonstrate similarly limited privacy preservation in this context. In contrast, the GenVC models show significant improvements over the other systems. Specifically, the GenVC-Small model achieves an EER of 29\%, while the GenVC-Large model achieves an EER of 27\%, outperforming the best VC-based approach (YourTTS) by over 15\%.

\begin{table}[th]
  \caption{Anonymization performance of VC-based approaches.}
  \label{tab:anon}
  \footnotesize
  \begin{center}
  \begin{sc}
  \vskip -0.1in
  \begin{tabular}{l cc}
    \toprule
    \textbf{System} & \makecell{\textbf{Privacy} \\ EER(\%) $\uparrow$ } & \makecell{\textbf{Utility} \\  WER(\%) $\downarrow$ } \\ 
    \midrule
    YourTTS &11.8 & 6.31 \\
    FreeVC & 9.1 & \textbf{2.73} \\
    Neural VC & 10.4 & 3.50 \\
    \midrule
    GenVC-Small & \textbf{29.0} & 6.03 \\
    GenVC-Large & 27.0 & 5.65 \\
    \bottomrule
  \end{tabular}
    \end{sc}
  \end{center}
  \vskip -0.1in
\end{table}

In terms of content preservation, FreeVC performs best among the evaluated VC approaches. The GenVC models achieve a WER of approximately 6\%, comparable to that of YourTTS, indicating sufficient preservation of linguistic content.

\subsection{Speaker-Content Disentanglement}
The preceding subjective evaluation and synthesized samples demonstrate that our model effectively converts source speech into target voices while preserving linguistic content. Nevertheless, speech signals inherently encode multiple attributes—such as speaker identity and environmental context—which our model may implicitly learn to disentangle during conversion. In this experiment, we focus specifically on the disentanglement of speaker identity. To assess the extent to which various model components preserve speaker-specific information, we construct speaker verification systems using features extracted from the trained phonetic tokenizer and the Perceiver encoder, treating these modules as fixed feature extractors.

The first system uses phonetic tokens as input features, with an embedding table that maps each token to a continuous vector representation. We employ the ECAPA-TDNN architecture~\cite{Desplanques2020ECAPATDNNEC} to train a speaker verification model using the VoxCeleb1 and VoxCeleb2 datasets, without any data augmentation. During evaluation, verification scores are computed using cosine similarity. As shown in Table~\ref{tab:spv}, this system yields an equal error rate (EER) close to 50\%, suggesting that the phonetic token representations retain minimal speaker-specific information. In contrast, \textit{System 2} follows the same training and evaluation protocol but utilizes the output of the Perceiver encoder from the GenVC-Small model as input features. This system achieves a markedly lower EER of 17.7\%, indicating that the Perceiver encoder retains a significant amount of speaker-related information. It is worth noting that the Perceiver encoder is trained in an unsupervised fashion on the LibriTTS dataset, which contains approximately 2,300 speakers. As a result, it may not serve as an ideal feature extractor for speakers in the VoxCeleb datasets.

\begin{table}[h]
  \caption{The performance of verification models on \textit{VoxCeleb1-O}.}
  \label{tab:spv}
  \footnotesize
  \begin{center}
  \begin{sc}
  \vskip -0.1in
  \begin{tabular}{cl cl}
    \toprule
    \textbf{ID} & \textbf{Source Feat.} & \textbf{Front-end + Back-end} & EER(\%) $\downarrow$ \\ 
    \midrule
    1& Phonetic-Token & ECAPA-TDNN + Cos & 48.5 \rule{0.49cm}{6pt} \\
    2&Perceiver-Feat. & ECAPA-TDNN + Cos & 17.7 \rule{0.18cm}{6pt} \\
    3&Perceiver-Mean & Cos & 34.7  \rule{0.35cm}{6pt} \\
    4&Perceiver-Mean & PLDA  & 12.2 \rule{0.12cm}{6pt} \\
    \bottomrule
  \end{tabular}
    \end{sc}
  \end{center}
  \vskip -0.1in
\end{table}

We further evaluate the speaker-discriminative capability of the Perceiver encoder by applying simple back-end verification methods. \textit{System 3} computes the mean of the Perceiver output across time and uses cosine similarity as the scoring metric. It achieves an EER of 34.7\% on the \textit{VoxCeleb1-O} test set. In \textit{System 4}, we replace cosine scoring with a probabilistic linear discriminant analysis (PLDA) backend trained on the VoxCeleb datasets using the same mean Perceiver features, which significantly improves performance to an EER of 12.2\%. These results provide additional evidence that the Perceiver encoder effectively captures speaker-specific information. 

Overall, our findings highlight the effectiveness of the proposed model in sufficiently disentangling speaker identity during training: while the content tokens extracted from source speech contain minimal speaker information, the Perceiver encoder captures most of the speaker-related characteristics.

\section{Limitations and Future Work}
\label{sec:dis}
A significant challenge in zero-shot VC remains the accurate cloning of speaker timbre, especially for unseen voices, alongside other speaker-specific characteristics such as accent, emotion, prosody, and recording environment~\cite{li2023freevc, wang2023lm}. These factors are essential for both achieving expected target voice conversions and effectively anonymizing source voices. Addressing this challenge requires access to more diverse training data, including recordings from varied acoustic environments~\cite{microvalle}. While such data may not match the quality of studio-recorded utterances, it could substantially improve the robustness of VC models in real-world applications. We believe that self-supervised learning approaches, such as GenVC, hold promise for enabling scalable training and enhancing generalization to unseen speakers and conditions.

However, our current study is subject to several limitations, mainly due to computational constraints. For instance, we did not explore the impact of model size on performance, despite scaling laws playing a critical role in optimizing LMs for large-scale training~\cite{kaplan2020scaling}. We did not investigate the influence of acoustic and phonetic token configurations or evaluate alternative SSL models, which could offer valuable insights compared to the ContentVec model used in our experiments. Addressing these gaps will be a key focus of future work. In addition, optimizing GenVC’s streaming capabilities and enhancing its robustness for multilingual and cross-lingual VC can be another interesting future work. 

\section{Conclusions}
\label{sec:conclue}
This paper introduces GenVC, a self-supervised zero-shot voice conversion system designed to leverage in-the-wild data for training. Unlike other LM-based approaches, GenVC adopts a simpler architecture and does not rely on external supervised models to disentangle linguistic content and speaker-specific attributes. Our experimental results show that GenVC delivers outstanding voice cloning capability while maintaining naturalness competitive with state-of-the-art zero-shot voice conversion systems. Moreover, GenVC effectively modifies the prosodic characteristics of the source speech, significantly enhancing privacy preservation in anonymization applications.

\section*{Acknowledgment}
This work was supported by the Office of the Director of National Intelligence (ODNI), Intelligence Advanced Research Projects Activity (IARPA), via the ARTS Program under contract D2023-2308110001. The views and conclusions contained herein are those of the authors and should not be interpreted as necessarily representing the official policies, either expressed or implied, of ODNI, IARPA, or the U.S. Government. The U.S. Government is authorized to reproduce and distribute reprints for governmental purposes notwithstanding any copyright annotation therein.

\bibliographystyle{IEEEtran}
\bibliography{ref}

\newpage
\appendix

\subsection{Impact Statement}
This paper presents work aimed at advancing the field of Voice Conversion. While our work holds significant promise, it also carries potential societal implications that warrant consideration. GenVC is a powerful voice conversion system capable of transforming source speech into desired voices. While this technology has valuable applications, such as enhancing privacy by anonymizing voices and enabling accessibility for individuals with speech impairments, it is also presents ethical challenges. Specifically, the ability to convincingly replicate voices can be misused to create audio deepfakes, which may be employed for malicious purposes, such as identity theft, fraud, and the spread of misinformation.

To mitigate these risks, we strongly advocate for the responsible and ethical use of voice conversion technologies. Researchers, developers, and users must comply with relevant laws and guidelines, ensuring that these systems are used exclusively for legitimate and beneficial applications. Transparency, informed consent, and robust safeguards should be prioritized to prevent misuse and protect individuals’ rights and privacy.

Furthermore, we emphasize the importance of continued research into developing countermeasures, such as deepfake detection tools, and fostering collaboration between researchers, policymakers, and industry stakeholders to address the ethical implications of voice conversion technology. By proactively addressing these concerns, we can maximize the benefits of systems like GenVC while minimizing their potential for harm. To support responsible use, we are releasing our code and models to the research community.

\subsection{Model Architecture}
\label{apx:mdlarc}

\subsubsection{DVAE}

Our DVAE encoder and decoder are both made up of a sequence of 1-D convolutional layers. The first two encoder convolutional layers upsample the input dimensionality to the DVAE hidden dimension of $1024$. They are followed by $3$ Resblocks \cite{he2016residual} and a final 1-D convolutional layer that projects the representation from the DVAE hidden dimension to the codebook dimension of $512$. Conversely, the decoder is a mirror image of the encoder, consisting the above-mentioned encoder components but in reverse order.

\subsubsection{Perceiver Encoder}
Our Perceiver encoder module comprises learned latent queries and a cross-attention mechanism. The learned queries attends to all input frames, transforming them into fixed-length representations. The encoder consists of four multi-head attention blocks, each with 8 heads, where each head has an inner dimension of 64.

\subsection{Subjective Evaluation Instructions}
\label{apx:subeva}
\subsubsection{Naturalness}
In this experiment, please listen to the speech sample and rate their naturalness on a scale from 1 (Bad) to 5 (Excellent), with increments of 0.5. 1, 1.5, 2, 2.5, 3, 3.5, 4, 4.5 and 5
are allowed on the half-point scale. The scale is defined as follows:

\begin{enumerate}[label*=\arabic*.]
    \item \textbf{Bad}: Very unnatural speech, completely unrecognizable as human speech.
    \item \textbf{Poor}: Noticeably unnatural speech with many artifacts and discontinuities.
    \item \textbf{Fair}: Moderately unnatural speech with noticeable artifacts, but not entirely unnatural.
    \item \textbf{Good}: Mostly natural speech with only minor artifacts that are noticeable upon careful listening.
    \item \textbf{Excellent}: Perfectly natural speech, indistinguishable from human speech.
\end{enumerate}

\subsubsection{Similarity}
In this experiment, you will listen to pairs of speech samples and rate how similar the second sample sounds to the reference speech in terms of speaker voice, speaking style, and environmental background, regardless of the content (which will differ).  Use a scale from 1 (Bad) to 5 (Excellent), with increments of 0.5. The allowed ratings are 1, 1.5, 2, 2.5, 3, 3.5, 4, 4.5, and 5. The scale is defined as follows:

\begin{enumerate}[label*=\arabic*.]
    \item \textbf{Bad}: The two samples sound completely different with barely any similarity.
    \item \textbf{Poor}: The two samples have some resemblance, but the differences are significant. For example, the speakers may differ in gender or pitch (e.g., a high-pitched female voice vs. a low-pitched male voice). Speaking style and environmental background also differ noticeably.
    \item \textbf{Fair}: The two samples sound somewhat similar, but there are still noticeable differences. It’s clear the speakers are the same gender, but their voices are distinctly different, and their speaking styles differ somewhat.
    \item \textbf{Good}: The given utterance sounds quite similar to the reference utterance, with only minor differences noticeable upon close listening. The speaker voices are close, and the speaking styles largely match.
    \item \textbf{Excellent}: The given utterance sounds identical to the reference utterance, with no perceivable differences. The timbre of the speakers is the same, their speaking styles match perfectly, and the environmental background is similar, including any acoustic noise.
\end{enumerate}

\end{document}